\documentclass[%
 reprint,
 amsmath,amssymb,
 aps,
]{revtex4-1}

\usepackage{graphicx}
\usepackage{dcolumn}
\usepackage{bm}
\usepackage{placeins}
\usepackage{xcolor}

\usepackage{hyperref}
\hypersetup{
    colorlinks=true,
    linkcolor=blue,
    filecolor=blue,      
    urlcolor=blue,
    citecolor=blue,
}

\usepackage{upgreek} 
\usepackage{comment}

\begin{document}

\preprint{APS/123-QED}

\title{Design of a prototype laser-plasma injector for the DESY-II synchrotron}

\author{S.~A.~Antipov\textsuperscript{1}}\email{sergey.antipov@desy.de}
\author{A.~Ferran~Pousa\textsuperscript{1}}
\author{I.~Agapov\textsuperscript{1}}
\author{R.~Brinkmann\textsuperscript{1}}
\author{A.~R.~Maier\textsuperscript{1}}
\author{S.~Jalas\textsuperscript{2}}
\author{L.~Jeppe\textsuperscript{2}}
\author{M.~Kirchen\textsuperscript{2}}
\author{W.~P.~Leemans\textsuperscript{1,3}}
\author{A.~Martinez de la Ossa\textsuperscript{1}}
\author{J.~Osterhoff\textsuperscript{1}}
\author{M.~Th\'evenet\textsuperscript{1}}
\author{P.~Winkler\textsuperscript{1,2}}

\affiliation{\textsuperscript{1}Deutsches Elektronen-Synchrotron DESY, Notkestr. 85, 22607 Hamburg, Germany}
\affiliation{\textsuperscript{2}Center for Free-Electron Laser Science, Luruper Chaussee 149, 22761 Hamburg, Germany}
\affiliation{\textsuperscript{3}Department of Physics  Universit\"at Hamburg, Luruper Chaussee 149, 22761 Hamburg, Germany}

\date{\today}

\begin{abstract}
The present state of progress in laser wakefield acceleration encourages considering it as a practical alternative to conventional particle accelerators. A promising application would be to use a laser-plasma accelerator as an injector for a synchrotron light source. 
Yet, the energy spread and jitter of the laser-plasma beam pose a significant difficulty for an efficient injection. In this paper we propose a design of a prototype injector to deliver 500 MeV low-intensity electron bunches to the DESY-II electron synchrotron. The design utilizes presently available conventional accelerator technology, such as a chicane and an X-band radio frequency cavity, to reduce the energy spread and jitter of the electron beam down to a sub-per-mille level.
\end{abstract}

\pacs{Valid PACS appear here}
\maketitle


\section{Introduction}
Using plasma as an accelerating medium has been attracting attention in the accelerator community for years, promising unmatched accelerating gradients and compact, energy-efficient acceleration~\cite{bib:Laser_acceleration}.  Laser-plasma accelerators (LPAs) in particular have shown a significant progress, achieving (although in different setups) GeV energies~\cite{bib:8GeV}, narrow energy spectrum~\cite{bib:Geddes}, and low emittance~\cite{Plateau:2012, Weingartner:2012kj}. Recently, the LUX LPA at DESY demonstrated a percent level energy stability during a 24-hour-long operation run~\cite{bib:LUX-24h}. While some challenges and limitations have been identified, in particular for collider applications ~\cite{bib:Whittum,bib:Lebedev,bib:Nagaitsev,Lehe:2017gry}, this progress encourages considering the LPA technology as a possible injector for various machines, including storage rings~\cite{bib:Hillenbrand}. 

In a synchrotron light source, an LPA injector could be used to top up the storage ring, significantly lowering the load on the conventional injector chain. Ultimately, by replacing the conventional injector the LPA would significantly reduce the spatial footprint and energetic cost of the accelerator complex. To achieve this, the LPA injector must deliver sufficient charge within the phase-space acceptance of the storage ring to compensate for the beam charge losses. 
As a reference, we consider the proposed 6~GeV PETRA~IV~\cite{bib:PIV} fourth-generation light source with a total beam charge up to 1600~nC and a lifetime of several hours, depending on the mode of operation. It is expected to feature a momentum acceptance of $1-2$\% and a transverse acceptance of $\sim 0.5$~$\upmu$m with realistic lattice errors. The light source could make use of an LPA-based top-up injector, delivering, e.g., $50-100$~pC bunches at $1-10$~Hz with an rms geometric emittance below 1~nm (12~$\upmu$m normalized) and an rms energy spread and jitter well below 1\%. While electron bunches with sufficient charge and low emittance are readily available from current LPA systems, achieving the required level of energy spread and jitter poses a significant challenge. 


This work proposes a design for a proof-of-concept low-energy (500~MeV) LPA injector capable of delivering the required beam energy bandwidth and stability to enable efficient injection into state-of-the-art storage rings. The design is based on the existing driving laser infrastructure at DESY~\cite{bib:LUX, bib:LUX-24h, bib:Opt_beam_loading, bib:LUX-optimization} and employs a beam energy compression and stabilization strategy~\cite{bib:Angel}, specifically conceived for the ultra-short and high-current LPA beams. The hereby proposed beamline is based solely on existing conventional accelerator technology to capture and transport the LPA beams and, finally, reduce their energy spread and jitter to the sub-per-mille level. This prototype could be built at DESY using existing equipment to enable injection into the DESY~II booster synchrotron~\cite{bib:PETRA-III_TDR} and serve as a testbed platform towards a final design at 6~GeV for PETRA IV. 
The design of the beamline is directly scalable to higher beam energies, upon the necessary technology development for the LPA to achieve current quality performance at the multi-GeV level.

\begin{figure*}[h!t]
   \centering
   \includegraphics[width=\textwidth]{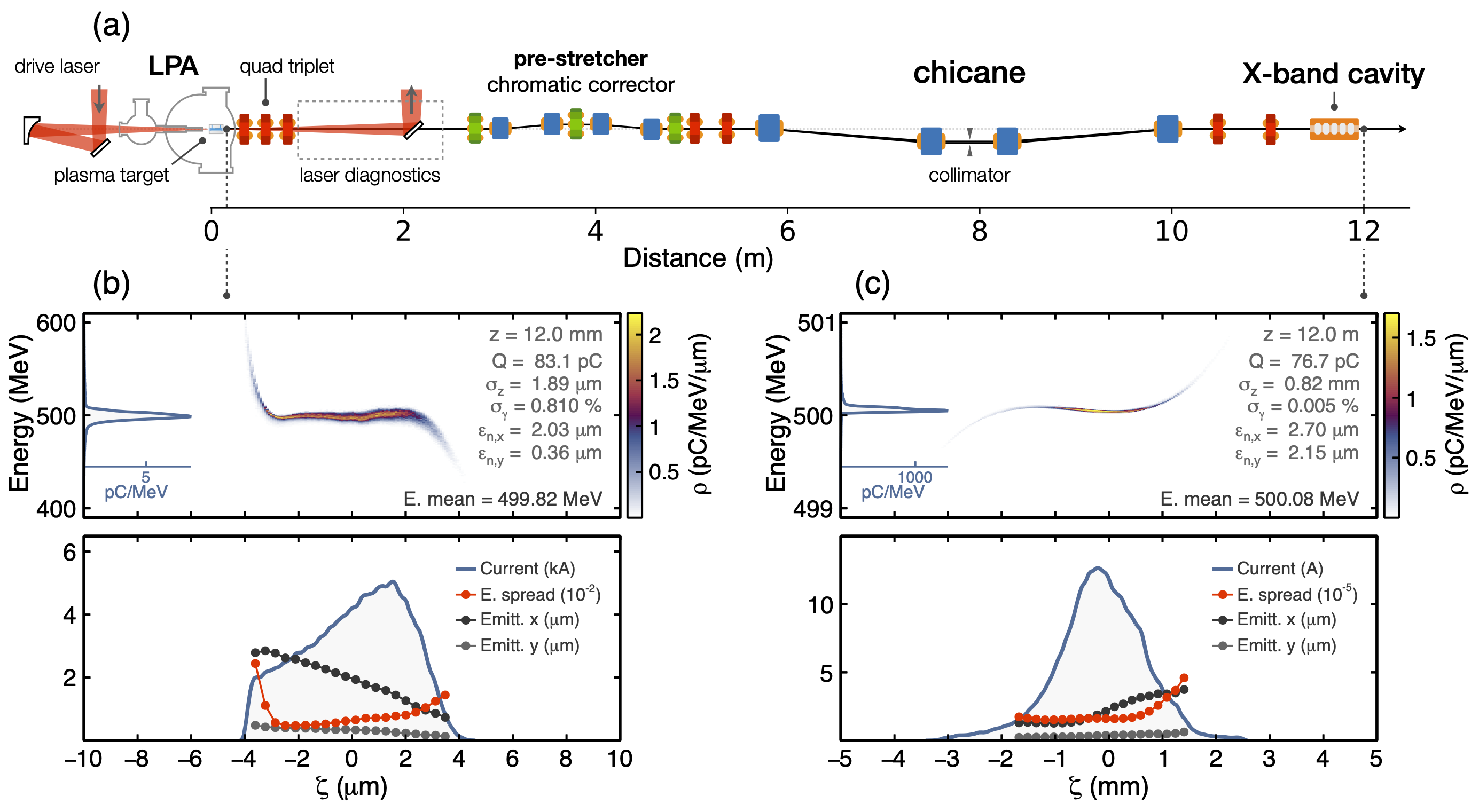}
   \caption{Schematic of the laser-plasma injector prototype (a): the laser-plasma accelerator  produces a 500 MeV beam with $\sim1\%$ energy spread; a quadrupole triplet (red) captures the beam and matches it to a dispersive section (chicane), made up of dipoles (blue) and sextupoles (green), that corrects the chromatic emittance growth. The beamline continues with a chicane that induces a large longitudinal decompression and a linear energy-time correlation over the beam. This energy correlation is canceled out by a short X-band rf cavity (orange), compressing the energy output of the beam by a factor 160. Electron bunch at the plasma cell exit, simulated with the particle-in-cell code FBPIC (b) and after the beamline, simulated with particle tracking code Ocelot (c): longitudinal phase-space distribution (top); current profile, slice energy spread, and normalized emittances (bottom). $\zeta = z - ct$ is the comoving coordinate of the bunch. The energy spread of the beam is computed via the median absolute deviation (see Sect.~\ref{sec:lpa}).}
   \label{fig:prototype_and_beams}
\end{figure*}

In this paper, we describe in detail the different components of the LPA injector prototype (Fig.~\ref{fig:prototype_and_beams}a), provide precise start-to-end simulations to assess the performance of the system, and discuss the scalability of the design to higher energies. 


\section{Laser-Plasma Accelerator}
\label{sec:lpa}
The configuration of the LPA is conceptually identical to that of the LUX experiment in Ref.~\cite{bib:Opt_beam_loading,bib:LUX-optimization}, where a specially tailored gas capillary target is used for controlling the injection and acceleration in the optimal beam-loading regime to yield electron beams with 1\% level energy spreads. This plasma target consists of two sections: first, a gas mixture of hydrogen and nitrogen forms a density spike where electron beam injection occurs, and second, a longer and uniform hydrogen gas section, which sustains the plasma wakefield for the acceleration of the beam up to the design energy $E_0 = 500$~MeV. After the acceleration, the electron beam is released through a tailored plasma density downramp to reduce its divergence and mitigate the chromatic emittance increase during the free drift towards the beamline. The drive laser considered is also identical to the LUX case~\cite{bib:Opt_beam_loading,bib:LUX-optimization}: the Ti:sapphire laser system ANGUS ($0.8~{\rm \upmu m}$ wavelength) with a pulse energy of 2.45~J, a pulse duration of $34~{\rm fs}$ (fwhm in intensity), and a peak power of 69~TW. The laser is focused into the plasma target by an off-axis parabolic mirror with a focal length of 2~m. It produces a 25~$\upmu$m spot size (fwhm in intensity), resulting in a peak normalized vector potential of $a_0 = 2.1$.

\begin{figure}[!b]
\centering
\includegraphics[width=\columnwidth]{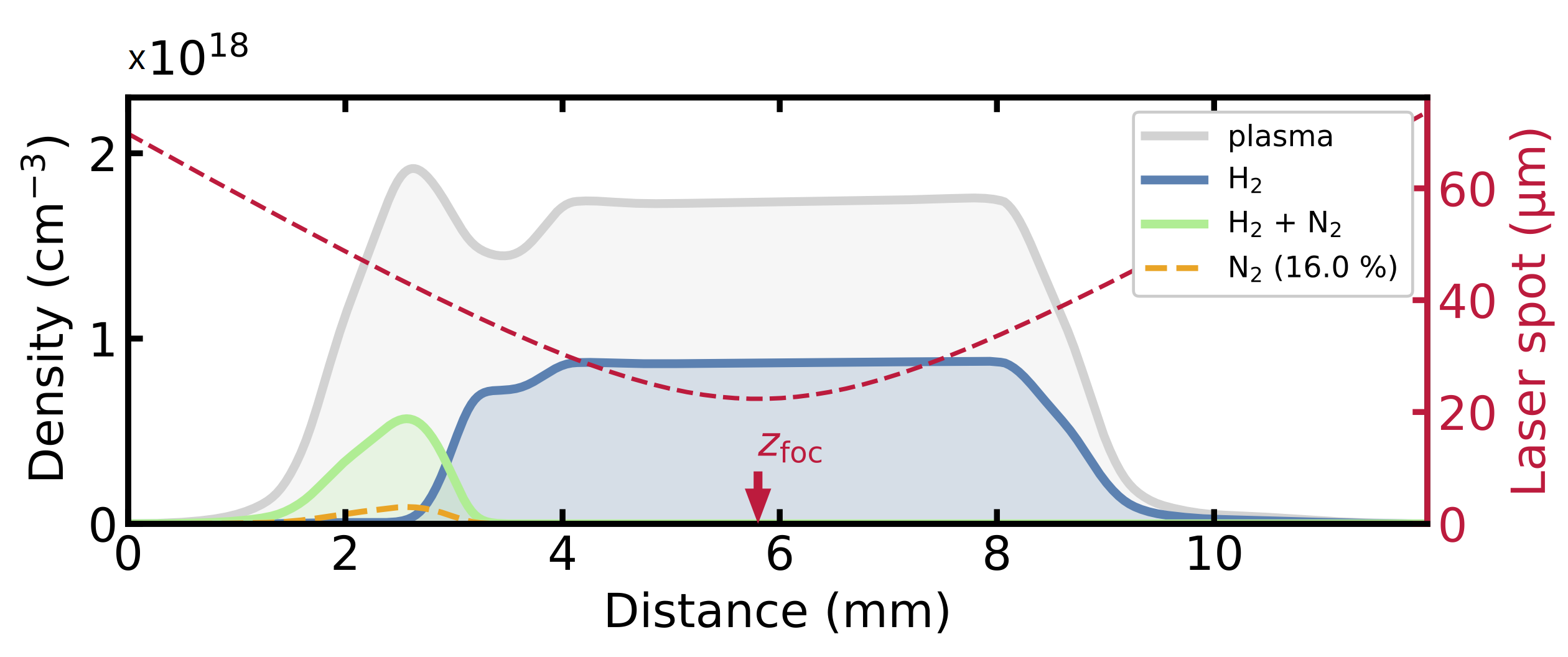}
\caption{Configuration of the plasma target providing optimized electrons beams at 500 MeV. The longitudinal distribution of the different gaseous species forming the target and the resulting plasma density are shown together with the laser spot size evolution in vacuum.}
\label{fig:plasma-profile}
\end{figure}

In order to produce suitable electron beams, the key parameters of the plasma target and the drive laser have been subjected to a Bayesian optimization procedure~\cite{bib:LUX-optimization} using the FBPIC particle-in-cell code~\cite{bib:FBPIC}: the overall plasma density value, the peak density of the first plasma spike, the nitrogen concentration, the length of the acceleration section, and the focal longitudinal position of the drive laser are scanned to find the point in parameter space which provides electron beams with a denser and narrower spectra peaking at 500 MeV. In our numerical simulations the laser pulse is modelled with a flattened Gaussian beam, which approximates the measured radial intensity evolution of the ANGUS laser used in LUX. 
Figure~\ref{fig:plasma-profile} summarizes the optimal configuration for the plasma target and the laser focal position. 
Figure~\ref{fig:prototype_and_beams}b presents the resulting electron beam distribution: the bunch has a charge of 83~pC, normalized emittances of 2.0 and 0.4~$\upmu$m (larger in the laser polarization plane), and an rms energy spread of 2.3\%, most of which comes from the distribution tails.  The bunch core (93\% of particles with the energy deviation within 3\%) has a negligibly small chirp and a relatively small effective energy spread of 0.8\%. To more conveniently characterize the bandwidth of the beam's spectrum, we adopt the following definition for the energy spread 
\begin{equation}
    \sigma_\gamma \equiv 1.48~\Delta_\gamma^{\rm mad},
    \label{eq:mad}
\end{equation}
with $\Delta_\gamma^{\rm mad}$ being the median absolute deviation of the energy distribution. This definition of the spread coincides with the rms value in case of a Gaussian distribution and provides a more accurate measure of the spectral bandwidth when the distribution differs from a Gaussian significantly. At the plasma cell exit, we obtain $\sigma_{\gamma,0} = 0.8\%$ for the simulated beam (Fig.~\ref{fig:prototype_and_beams}b). 
As shown in~\cite{bib:Opt_beam_loading}, this working point for the LPA is mostly affected by fluctuations in the laser focal position, which alters the amount of injected charge, and thus, the operation at optimal beam loading conditions. The potential effect of these fluctuations is considered in detail in Sect.~\ref{sec:jitter}. 

\section{energy compression scheme}
In order to be suitable for efficient injection into a light source the small energy spread and potential energy jitter of the LPA beam has to be further lowered by about an order of magnitude. This could be achieved in several ways: using a chicane and an active dechirper such as a radio frequency (rf) cavity or an active plasma cell; using a passive dechirper: corrugated, dielectric or plasma dechirper; or using an emittance exchange scheme. The potential gain of the emittance exchange~\cite{bib:EX} scheme seems to be limited, taking into account that the transverse and longitudinal emittances of the LPA beam are of a similar order of magnitude. Moreover, a special attention would have to be given to chromatic effects in order not to spoil the transverse emittance during the capture and transport. Reducing the energy spread with passive dechirpers would require stretching the bunch with a complex $R_{56}<0$ chicane, similar to the one presented in~\cite{bib:negative_r56}, so that the low energy particles move to the head and the high energy to the tail of the bunch. Such an approach might have potential problems with higher order dispersion and chromatic emittance increase. Moreover, the efficiency of energy compression would be affected by fluctuations of bunch charge. 

From the methods discussed above only active energy compression schemes have the crucial advantage of both reducing the energy spread and correcting the central energy. Such a beam energy compression and stabilization scheme has been recently proposed for laser-plasma accelerators~\cite{bib:Angel}. The method requires a simple chicane to stretch the bunch and create a linear correlation between particles' energies and their longitudinal positions (chirp) and an active dechirper that applies a linear kick to put the particles precisely on the design energy. Since the ultra short bunch lengths and high peak currents, available by LPA beams, are not required for applications in synchrotron light sources, it is favorable to opt for a large beam decompression and an rf dechirper. A combination of a chicane and an rf cavity is known to be an efficient way of energy compression as demonstrated, i.e. in~\cite{bib:Glasgow}. The X-band rf technology offers sufficiently short wavelengths and high gradients, both of which are beneficial for a dechirper. The X-band rf is mature and is widely used, for example in transverse deflective structures~\cite{bib:X-Band-TDS}. Following Ref.~\cite{bib:X-Band-CLIC} we can assume a safe operating accelerating gradient of 60~MV/m at the frequency $f_{\rm rf} = 12$~GHz. 
Provided the initial bunch length $\sigma_{z,0}$ is small, $\sigma_{z,0} \ll R_{56}\sigma_{\gamma,0}$, the required rf voltage $V_{\rm rf}$ can be found as
\begin{equation}
    V_{\rm rf} = E_0 c / q_e \omega_{\rm rf} R_{56},
    \label{eq:RF_Volt}
\end{equation}
where $\omega_{\rm rf} = 2\pi f_{\rm rf}$, $q_e$ the elementary charge, and $c$ is the speed of light. Finally, for the rf kick to be linear the bunch has to be short compared to the rf wavelength: $R_{56}\sigma_{\gamma,0} \ll c/f_{\rm rf}$. In practice, the nonlinearity of the rf kick will generate a small spread in the final beam energy, as discussed in Sect.~\ref{sec:final_spread}.

\section{Beamline design}
\begin{figure}[!b]
   \centering
   \includegraphics*[width=3.5in]{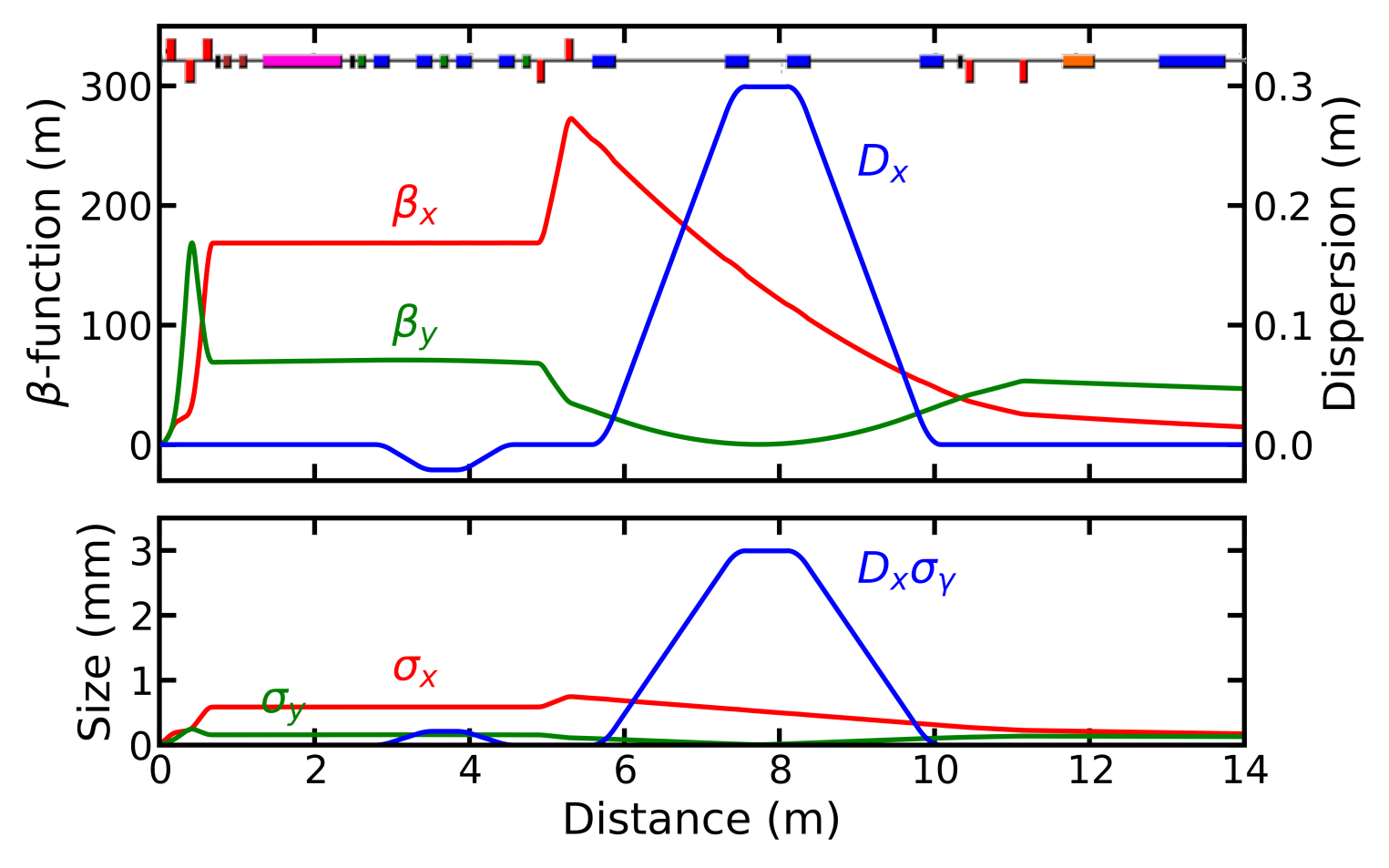}
   \caption{Beamline optics function (top) and rms beam sizes of a Gaussian beam (bottom). Dipoles are shown in blue, quadrupoles -- red, sextupoles -- green, X-band rf -- orange, BPMs -- black, a section for laser beam removal and diagnostics -- pink. The last magnet, a Y-chamber dipole of the present electron beamline, is turned off during the injection from the LPA. $\sigma_{x,y}$ denote the beam sizes in the limit $\sigma_\gamma \rightarrow 0.$}
   \label{fig:overall}
\end{figure}

Figure~\ref{fig:overall}~(top) depicts a schematic view of the beamline and its linear lattice functions computed by the Optim code~\cite{bib:optim}. First, the beam is captured by a quadrupole triplet, following the positive practical experience with beam capture with electromagnetic quadrupoles at LUX. We are assuming a bore aperture of 22~mm and the maximum gradient of 80~T/m. At this strength the transverse rms beam size in the triplet stays within 1~mm, preventing beam losses during the capture~[Fig.~\ref{fig:overall} (bottom)].
After the triplet follows a drift section for laser beam removal and diagnostics. Then a chicane with well separated $\beta$-functions reduces the chromatic emittance increase in the plane with higher emittance. It employs three sextupole magnets: one in the center of the chicane and two with half the strength and the opposite polarity on each side where dispersion $D = 0$ for reduction of geometric aberrations. Then follows the main chicane with $R_{56} = 10$~cm that stretches the electron bunch to 0.8~mm rms. The chicane creates an energy chirp of 5~MeV/mm. And finally, an X-band rf dechirper corrects the energy deviation. 
According to Eq.~(\ref{eq:RF_Volt}) a 20~MV voltage is sufficient for efficiently compensating up to 3\% energy deviation with the particles staying within the linear region of the rf slope. Particles outside this acceptance range can be efficiently disposed off by a collimator in the middle of the main chicane. Thanks to the large dispersion the particle's transverse position is dictated by its energy offset: $D_x\sigma_\gamma \gg \sigma_x = \sqrt{\epsilon_x \beta_x}$ [Fig.~\ref{fig:overall} (bottom)]. 

After the dechirper the beamline merges with the existing electron beamline from Linac to DESY-II synchrotron, where a present large-aperture dipole with a Y-chamber can be used to send the beam to DESY-II. Matching of the optics function is achieved using four relatively weak quadrupole magnets. Table~\ref{tab:beam_par} lists the beam parameters at the entrance and the exit of the beamline and Table~\ref{tab:sum} summarizes the key components.

\begin{table}[h]
   \centering
   \caption{Beam parameters at the entrance and at the exit of the beamline.}
   \begin{tabular}{lrr}
       \toprule
       \textbf{Parameter} & \textbf{Plasma cell exit}& \textbf{Final}\\
       \hline
           Twiss $\alpha$ (x,y) & -0.47, -0.42 & 1.35, 1.01\\ 
           Twiss $\beta$ (x,y) & 3.1, 3.0~mm & 14.5, 48.4~m\\
           Norm. $\epsilon$ (x,y) & 2.0, 0.4~$\upmu$m & 2.7, 2.1~$\upmu$m\\
           Charge & 83~pC & 77~pC\\
           Length $\sigma_z$ & 2.0~$\upmu$m & 0.8~mm\\
           Energy spread $\sigma_\gamma$ & $0.8\times 10^{-2}$ & $0.5\times 10^{-4}$\\
        \hline
   \end{tabular}
   \label{tab:beam_par}
\end{table}
\begin{table}[!h]
   \centering
   \caption{Parameters of key beamline elements.}
   \begin{tabular}{llcr}
       \toprule
       \textbf{Element} & \textbf{N} & \textbf{Length} & \textbf{Strength} \\
       \hline
           Capture Quad & 3 & 10~cm & $\leq 80$~T/m \\ 
           Main chicane Dip & 4 & 30~cm & $1$~T \\ 
           12~GHz Cavity & 1 &  40~cm & $< 60$~MV/m \\
           Chrom chicane Dip & 4 & 20~cm & $0.3$~T\\
           Sextupoles & 3 & 10~cm  & 1600~T/m$^2$ (cent.)\\
           Quadrupoles & 4 & 10~cm & $\leq 15$~T/m \\
       \hline
   \end{tabular}
   \label{tab:sum}
\end{table}
At 500~MeV the space charge (SC) interaction is already small with the SC parameter (SC tune shift per unit length) being $\delta Q_{\rm SC} \sim 0.02$~m$^{-1}$ at the exit of the LPA. Thus, it is not expected to play a significant role as the beam diverges after leaving the plasma cell, except perhaps in the first few centimeters immediately after the exit. The SC effect on beam sizes therefore should be negligible, as observed at LUX. Another physical effect that might change the dynamics is the coherent synchrotron radiation (CSR). Its characteristic scale, the energy change per unit length, can be estimated as~\cite{bib:CSR}:
\begin{equation}
    W_{\rm CSR} = N_b r_e m_e c^2 \frac{(\kappa\sigma_z)^{2/3}}{\sigma_z^2},
    \label{eq:CSR}
\end{equation}
where $N_b$ is the bunch population, $r_e$ -- classical electron radius, $m_e$ -- electron mass, $\kappa$ -- the curvature of the bending field, and $\sigma_z$ -- the bunch length. Due to the steep dependence on $\sigma_z$ it is beneficial to lengthen the bunch before it reaches the first strong dipole of the main chicane. This is done in the first chromatic chicane, which stretches the bunch from $\sigma_z = 2~\upmu$m to $\sigma_z \sim 30~\upmu$m, making the CSR in the main chicane negligible.

\section{Particle tracking}

In order to confirm the beamline performance we performed a series of macroparticle simulations in the Ocelot tracking code~\cite{bib:ocelot} using realistic LPA beam from the PIC simulation. The beam was modelled as an ensemble of $2\times10^5$ macroparticles of variable charge. Its distribution is presented in Fig.~\ref{fig:prototype_and_beams}b. Lattice elements: quadrupoles, dipoles, sextupoles, and the rf cavity were modelled by linear and second order maps with the transverse focusing of the rf taken into account. The code treats the CSR effect in a `projected' 1D model~\cite{bib:Saldin}, which calculates the longitudinal wake-field of the beam by projecting the real 3D beam distribution onto the reference trajectory.

\subsection{Emittance preservation}
\begin{figure}[!h]
   \centering
   \includegraphics*[width=0.90\columnwidth]{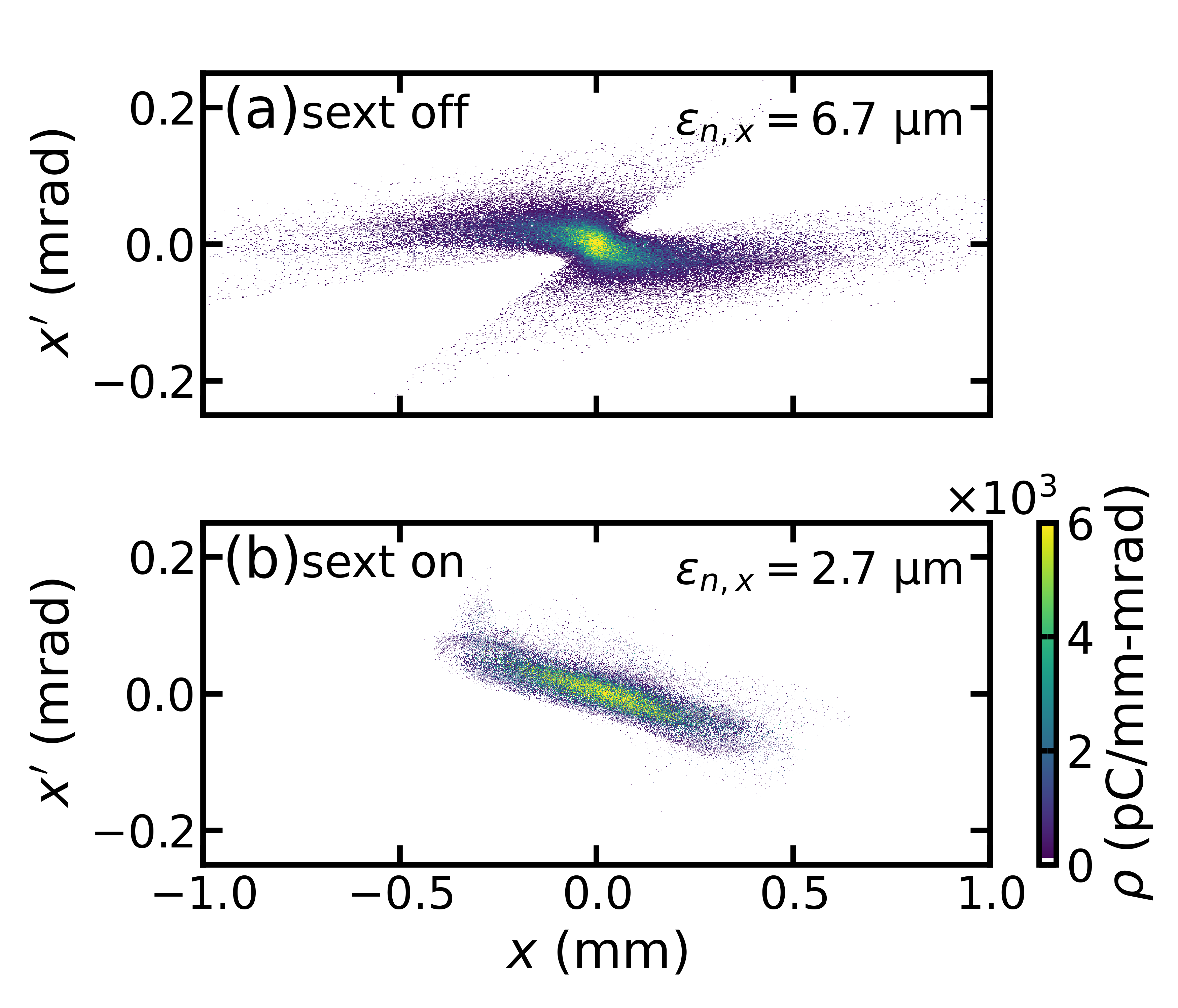}
   \caption{Using sextupoles allows significantly reducing the unwanted chromatic emittance increase in the horizontal plane. Bunch phase space densities at the exit of the beamline with the chromatic sextupoles off and on.}
   \label{fig:sex_on_off}
\end{figure}

Due to chromatic effects in the capture triplet the electron beam might suffer a significant chromatic emittance increase. The emittance growth is induced by the build-up of correlation between the transverse positions of the particles and their momenta, as shown in Ref.~\cite{bib:Floettmann}. 
Without correction, for the parameters of the considered LPA beam the chromatic emittance growth is rather significant with the final normalized emittance reaching 6.7~$\upmu$m in the horizontal plane, a degradation of more than three times the initial value (Fig.~\ref{fig:sex_on_off}a).
On the other hand, since the chromatic emittance increase is correlated, it can be reversed using sextupoles in a dispersive region. With the sextupoles on, the final normalized emittance is reduced to 2.7~$\upmu$m~(Fig.~\ref{fig:sex_on_off}b) with only a small degradation in the vertical plane. It shall be noted that the C-chicane presented here does not feature a complete cancellation of nonlinear aberrations from the sextupoles. A longer S-chicane with four sextupole magnets could likely offer some improvement.
\newpage
\subsection{Final energy spread and variation}\label{sec:final_spread}
Tracking simulations with a realistic beam distribution show a final projected energy spread of 25~keV, or $\sigma_\gamma = 0.5\times 10^{-4}$ in relative terms~(Fig.~\ref{fig:prototype_and_beams}c). Such a low value is possible thanks to a combination of a large bunch lengthening and a high linearity of the rf kick. 
Due to the large bunch lengthening in the chicane, the finite initial bunch length $\sigma_{z,0}$ only adds a minor spread to the slice energy of the beam: $\sigma_\gamma^{\rm fl} = \sigma_{z,0}/R_{56} \sim 2\times 10^{-5}$~(Fig.~\ref{fig:prototype_and_beams}c).
With $\sigma_\gamma^{\rm fl}$ being small the final projected energy spread is essentially governed by the nonlinearity of the rf kick:
\begin{equation}
    V(t) = V_{\rm rf} \sin (\phi) = V_{\rm rf}\, [\phi - \frac{1}{6}\phi^3 + \mathcal{O}(\phi^5)],
    \label{eq:nonl_rf}
\end{equation}
where $\phi = \omega_{\rm rf} t$ is a time-dependent rf phase. The second term in the RHS of Eq.~(\ref{eq:nonl_rf}) ($\propto \phi^3$) adds a nonlinear term to the ideal linear kick $V_0 = V_{\rm rf}\phi$: particles with lower energies will receive a smaller acceleration than needed, while those with higher energies -- a larger than required. This effect increases the final energy spread by (rms)
\begin{equation}
    \sigma_\gamma^{\rm nonl} = \frac{V_{\rm rf}q_e}{E_0} \frac{\sqrt{\langle\phi^6\rangle}}{6},
    \label{eq:sigma_nonl}
\end{equation}
where the averaging $\langle...\rangle$ goes over the whole ensemble of particles. Using Eq.~(\ref{eq:RF_Volt}) and a normal distribution,  Eq.~(\ref{eq:sigma_nonl}) becomes $\sigma_\gamma^{\rm nonl} \approx 0.645\,(\omega_{\rm rf} R_{56} /c)^2 \sigma_{\gamma,0}^3$. 
For the considered parameters we obtain $\sigma_\gamma^{\rm nonl} \approx 2\times 10^{-4}$.
The impact of the nonlinearity of the rf signal on the projected energy spread can be minimized by slightly increasing the voltage, by about 2\% with respect to the value given in Eq.~(\ref{eq:RF_Volt}). This has been done in the tracking simulations ~(Fig.~\ref{fig:prototype_and_beams}c), where the obtained total energy spread (computed via the median absolute deviation of the simulated spectra) is four times smaller than the previous estimate.

Another minor contribution to $\sigma_\gamma$ comes from the phase and voltage stability in the rf cavity. Contemporary off-the-shelf systems feature a control of amplitude and phase at the level of $10^{-3}$ and $0.1^{\circ}$~\cite{bib:LLRF}, while the beam loading effects in the rf can be safely neglected due to negligible average and peak currents. This allows assuming that the rf voltage seen by the bunch stays within $10^{-3}$ of the design value, adding $\sigma_\gamma^{\rm rf} \sim 10^{-5}$ to the final energy spread, so this effect was safely neglected in our tracking simulations.

Additionally, the final beam energy can be affected by effects that vary shot-to-shot, such as a timing jitter between the laser and the rf system or a jitter of the central beam energy. This shot-to-shot variation can, in principle, be larger than the single-shot energy spread. Assuming a Gaussian timing jitter with an rms $\sigma_t^{\rm tj}$ the resulting rms energy variation is 
\begin{equation}
    \sigma_\gamma^{\rm tj} = \frac{V_{\rm rf}q_e}{E_0} \omega_{\rm rf} \sigma_t^{\rm tj}.
    \label{eq:sigma_tj_1}
\end{equation}
Plugging in $V_{\rm rf}$ from Eq.~(\ref{eq:RF_Volt}) the variation becomes
\begin{equation}
    \sigma_\gamma^{\rm tj} =  c\, \sigma_t^{\rm tj}/R_{56}.
    \label{eq:sigma_tj_2}
\end{equation}
For a 100~fs rms jitter -- a level that can be realistically achieved nowadays -- one gets $\sigma_\gamma^{\rm tj} \approx 3 \times 10^{-4}$. This value is comparable with the contribution of the rf kick nonlinearity, and thus the effect should not be neglected.

Small variations of the initial central energy will also affect the final energy distribution.
As the central energy mismatch displaces the bunch center from the zero crossing of the rf, Eq.~(\ref{eq:RF_Volt}), it reduces the linearity of the rf kick. Thus, the jitter of the central energy generates a variation of the final beam energy through the nonlinearity of the rf kick. This effect can be estimated using Eq.~(\ref{eq:sigma_nonl}): for a 1\% rms Gaussian central energy jitter, neglecting the initial energy spread ($\sigma_{\gamma,0} \rightarrow 0$) and energy collimation in the chicane, one obtains the final energy variation $\sigma_\gamma^{\rm ej} \approx 4\times 10^{-4}$.


\subsection{Effect of initial energy jitter and chirp}\label{sec:jitter}
\begin{figure*}[!ht]
   \centering
   \includegraphics[width=7.0in]{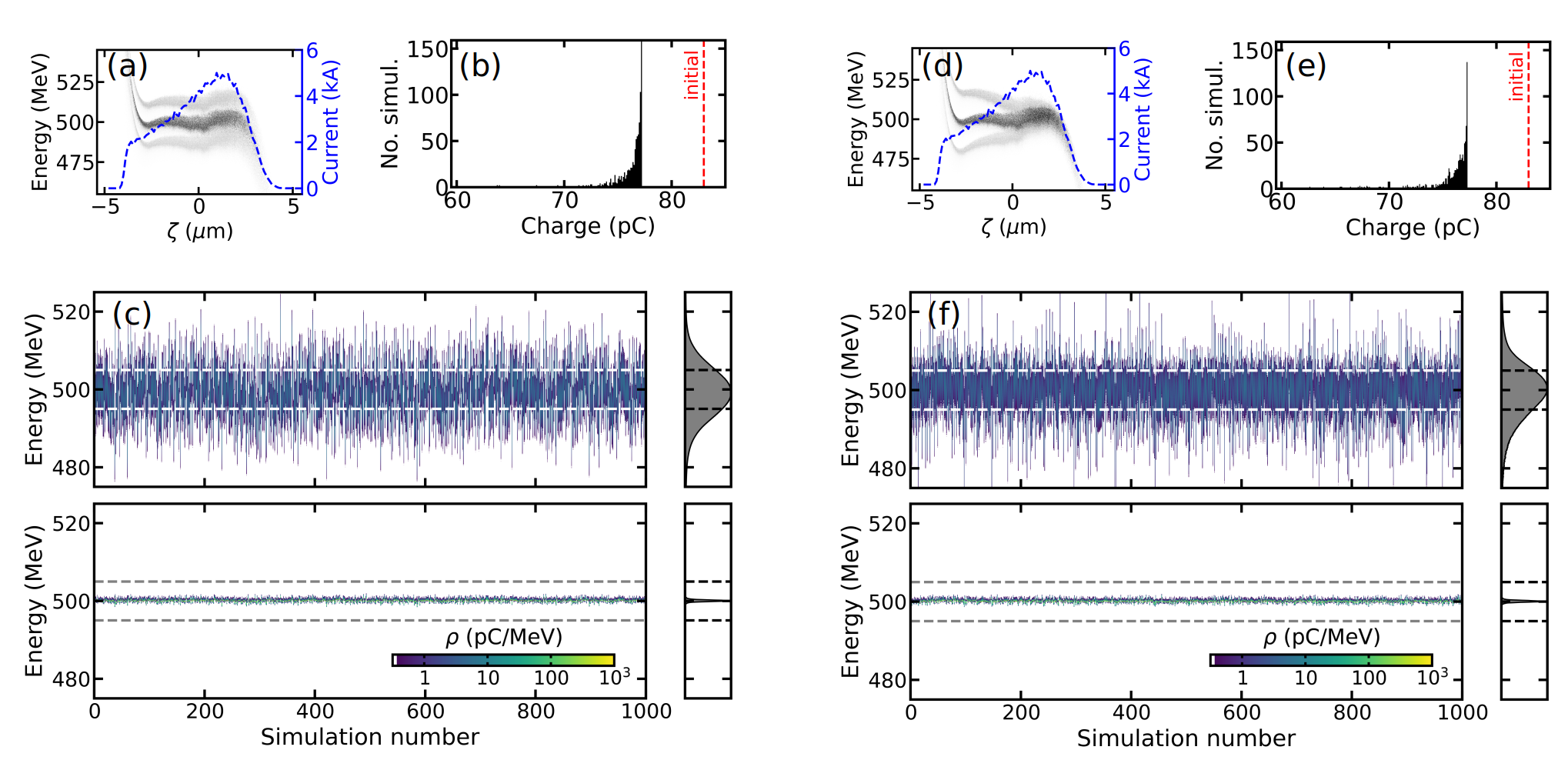}
   \caption{The active energy compensation scheme with an X-band dechirper efficiently reduces the energy spread of the LPA beam in the presence of energy jitter and chirp. Examples of initial longitudinal distributions when varying the central energy (a). Panel (b) shows the distribution of the bunch charge at the exit of the injector for 1000 simulated LPA beams with and rms central energy jitter of 1\%; the initial bunch charge of 83~pC is shown in red. (c) -- energy spectra at the exit of the plasma cell (top) and at the exit of the injector, after the energy compression (bottom); panels on the right show the average beam spectra. The dashed lines represent $\pm1$\% deviation from the design energy of 500~MeV. Panels (d,e,f) show the same results for a Gaussian spread of initial energy chirp. All simulation results include a Gaussian timing jitter between the LPA and the rf with an rms spread of 100~fs.}
    \label{fig:jitter_comp}
\end{figure*}
In a real world, the mean energy and chirp of the LPA beam will vary, for example, due to variations of the bunch charge resulting in different levels of beam loading~\cite{bib:Opt_beam_loading}. Recent experiments at LUX~\cite{bib:LUX-24h,bib:Opt_beam_loading,bib:LUX-optimization} give confidence that, with moderate improvements to the laser stability, a central energy stability of 1\% rms can be expected. In order to assess the injector's performance under such conditions, we performed a series of tracking simulations varying the central energy of the bunch, while keeping the remaining parameters constant. We assumed a normal distribution of the central energies with an rms spread of 1\%. Thus, the average electron energy at the exit of the LPA was $499.3\pm6.6$~MeV, with the variation computed according to the definition of Eq.~({\ref{eq:mad}). On top of that, we also assumed a Gaussian timing jitter between the LPA drive laser and the rf of $\sigma_t^{\rm tj} = 100$~fs rms.

After passing through the active dechirper the energy spread decreased, as expected. The average value and variation are $500.1\pm0.2$~MeV, which corresponds to a relative stability at the $4\times 10^{-4}$ level (Fig.~\ref{fig:jitter_comp}c). This result is consistent with our estimate of energy spread caused by the nonlinearity of the rf potential, Eq.~(\ref{eq:sigma_nonl}), after accounting for the collimation at $\pm 3$\% in energy.
The energy stabilization comes at the cost of the reduced bunch charge. Particles with too large energy offsets that cannot be efficiently captured and transported are stopped by the collimator. This results in a slight shot-to-shot variation of the bunch charge: $75.5 \pm 4.3$~pC~(Fig.~\ref{fig:jitter_comp}b). The effect of the timing jitter was found to be negligible: with the jitter excluded from the simulation both the final energy spread and the final charge distributions were found to be unaltered. 

If the beam loading throughout the plasma acceleration process is sub-optimal, the energy chirp of the bunch might not be fully cancelled at the exit of the plasma. Then the head of the bunch that sees no self-wake will remain at the designed energy while the tail will have a higher or a lower energy depending on the sign of the chirp. Figure~\ref{fig:jitter_comp}d presents examples of such distributions. They were obtained by fixing the energy of the head, defined at a point were the bunch current reaches $1/e$ of its peak value, and applying a linear chirp to the rest of the beam. These model beams resemble qualitatively the experimental observations at LUX~\cite{bib:Opt_beam_loading,bib:LUX-optimization}.

Assuming the initial normal distribution of chirps corresponding to a 1\% rms spread of central energies (the median electron energy and its variation in this case are  $499.6\pm6.6$~MeV) one obtains the final energy variation of $500.1\pm0.2$~MeV (Fig.~\ref{fig:jitter_comp}f). The final charge in this case varies slightly from the case of naive shifting the central energy: $74.5 \pm 5.7$~pC (Fig.~\ref{fig:jitter_comp}e). Given that our goal is to provide a top-up of a small fraction of the beam, a sub-ten percent charge variation seems acceptable.

In the present simulation setup we assumed the energy jitter levels that are achievable with today's laser-plasma accelerators, which run at low repetition rate (1~Hz) and have no dedicated feedback stabilization of the injected charge and central energy. With the advent of higher repetition rate (10~Hz -- 1~kHz) high-power lasers and fast feedback systems we expect these jitters to further reduce. Thus our results for the final energy spread and bunch charge variation might be too conservative. It should be noted though that improving the quality and stability of LPA beams one might at some point become limited by the timing jitter.
 
\section{Scalability to higher energies}

The presented design is readily scalable to a higher 6~GeV energy that would allow a direct injection into the PETRA~IV~\cite{bib:PIV} light source. Preliminary studies provide the following estimated beam parameters at the LPA exit: $Q = 100$~pC, $\sigma_{\gamma,0} = 10^{-2}$, rms divergence $<0.2$~mrad, $\sigma_{z,0} \sim 4~\upmu$m. This beam can be captured using a triplet of quadrupoles similar to ESRF-EBS with a field gradient of 100~T/m, a length of 50~cm, and a bore radius of 12.5~mm~\cite{bib:ESRF-EBS}. A potential issue could arise from CSR in the main chicane dipoles -- a steady state estimate yields a wake of 1.5~MV/m for a 100~pC bunch. This can be mitigated by a slight bunch stretching in the first, weaker chicane. Preliminary studies show that the bunch stretched to 50~$\upmu$m~rms length does not experience significant CSR effects: The steady-state estimate, Eq.~(\ref{eq:CSR}), yields $W_{\rm CSR} < 1$~MV/m for 1.5~T dipoles of the main chicane. This stretching can be achieved with a relatively compact, 3-m-long, chicane. The total rf required voltage would be about 300~MV, or 5\% of what one would need to be provided by a conventional linac. Given the high gradients achievable in state-of-the-art X-Band structures, the total length of the rf section could be within 5~m. The complete injector, including laser in- and out-coupling can be within 50~m. For comparison, a 4~GeV LCLS-II superconducting electron linac spans over 1~km~\cite{bib:LCLS}, a 17~GeV linac of the European XFEL occupies 1.6~km \cite{bib:XFEL}, a 300~MeV IOTA FAST linac -- about 40~m \cite{bib:IOTA_FAST}, and a proposed 1~GeV warm X-band linac of the EuPRAXIA SPARC\_LAB project is about 50~m long~\cite{bib:EuPRAXIA}.

\section{Conclusion}

LPAs have the potential to offer a compact, cost-effective alternative to conventional accelerators to serve as injectors for future light sources. One of the challenges on this way is the large spread and jitter of the beam energy, which would significantly limit the injection efficiency. In this paper we have considered a solution to the problem of capturing, transporting, and compressing the energy output of LPA beams solely by means of conventional accelerator technology. Building up on the positive experimental experience at LUX, we conclude that the capture of a beam with an rms energy spread up to 1\% is feasible without unacceptable chromatic emittance degradation for the simulated beam distributions. The chromatic emittance increase occurring during the capture can be efficiently minimized using a combination of a C-chicane and sextupole magnets. After that, the energy output can be drastically reduced, reaching sub-per-mille levels, using a compact X-band rf dechirper. The dechirper requires only a small fraction ($\leq 5$\%) of the rf voltage, needed by a conventional linac.

To demonstrate the fruitfulness of this approach we have designed a 500~MeV prototype injector for the DESY-II synchrotron. It would deliver low-charge, $Q \sim 80$~pC, tests beams to the synchrotron at a repetition rate up to 1~Hz. The prototype could operate in parallel to the existing conventional electron linac and share the already existing injection infrastructure. Tracking simulations show that with the chromaticity correction the transverse emittance can be kept at a competitive level of $2~\upmu$m, while the bunch energy spread can be reduced to $10^{-4}$ level. Considering a conservative estimate of the LPA central energy jitter and chirp as well as a realistic timing jitter between the LPA and the rf, we have confirmed the robustness of the proposed approach: the charge losses do not exceed $\sim 10$\% on average, while the beam remains centered at the desired energy: $500.1\pm0.2$~MeV.

The design of the injector provides scalability to higher energies. In particular, it can be extended up to 6~GeV, the beam energy of the PETRA~IV light source, pending the technical development of an LPA that could deliver electron beams at these energies with the presently achieved levels of beam quality and energy stability. 

\section*{Acknowledgements}
We would like to thank Sergey Tomin for his help with the Ocelot code and Holger Schlarb for clarification on the rf-to-optical timing jitters. 
This research was supported in part through the Maxwell computational resources operated at Deutsches Elektronen-Synchrotron DESY, Hamburg, Germany

\end{document}